\documentclass[12pt]{article}

\begin{document}

\title{Geometry as an object of experience: \\Kant and the missed debate between \\Poincar\'{e} and Einstein}

\author{Shahen Hacyan}

\renewcommand{\theequation}{\arabic{section}.\arabic{equation}}

\maketitle
\begin{center}

{\it  Instituto de F\'{\i}sica,} {\it Universidad Nacional Aut\'onoma
de M\'exico,}

{\it Apdo. Postal 20-364, M\'exico D. F. 01000, Mexico.}

e-mail: hacyan@fisica.unam.mx

\end{center}

%\vskip0.5cm

\vskip0.5cm

\abstract Poincar\'{e} held the view that geometry is a convention
and cannot be tested experimentally. This position was apparently
refuted by the general theory of relativity and the successful
confirmation of its predictions; unfortunately, Poincar\'{e} did not
live to defend his thesis. In this paper, I argue that: 1) Contrary
to what many authors have claimed, non-euclidean geometries do not
rule out Kant's thesis that space is a form of intuition given {\it a
priori}; on the contrary, Euclidean geometry is the condition for the
possibility of any more general geometry. 2) The conception of
space-time as a Riemannian manifold is an extremely ingenious way to
describe the gravitational field, but, as shown by Utiyama in 1956,
general relativity is actually the gauge theory associated to the
Lorentz group. Utiyama's approach does not rely on the assumption
that space-time is curved, though the equations of the gauge theory
are identical to those of general relativity. Thus, following
Poincar\'{e}, it can be claimed that it is only a matter of
convention to describe the gravitational field as a Riemannian
manifold or as a gauge field in Euclidean space.

\vskip1cm

\begin{center}
*~~~~~*~~~~~*~~~~~*~~~~~*

\end{center}

 \vskip1cm

\endcenter

Many scientists and philosophers have argued that the possibility of
non-Euclidean geometries contradicts Kant's thesis that geometric
axioms are synthetic judgments {\it a priori} (Kant 1929, 1986).
Accordingly, it became a common place to assert that the general
theory of relativity, by revealing the non-Euclidean nature of
physical space, refuted Kant's doctrine on the transcendental nature
of space. However, Poincar\'{e} (1952) sustained that geometry is a
convention and cannot be the object of any experience, but his point
of view fall into oblivion due to the commonly held belief that the
Riemannian nature of space-time is testable.

The plan of the present paper is as follows: In Section 1, a brief
historical introduction to the problem is given and then it is shown
that the existence of Riemannian geometry not only does not refute,
but confirms Kant's doctrine. In Section 2, the geometric
conventionalism of Poincar\'{e} is reviewed. Section 3 is devoted to
the alternative formulation of general relativity, as a gauge theory,
given by Utiyama (1956). Finally, Section 4 compares the ideas of
Poincar\'{e} and Einstein on the physical nature of space following a
text written by the latter in 1949.

\section{How can Riemannian geometry be possible?}

The argument that non-Euclidean geometries contradict Kant's doctrine
on the nature of space apparently goes back to Helmholtz (1995) and
was retaken by several philosophers of science such as Reichenbach
(1958) who devoted much work to this subject.

In a essay written in 1870, Helmholtz (1995) argued that the axioms
of geometry are not {\it a priori} synthetic judgments (in the sense
given by Kant), since they can be subjected to experiments. Given
that Euclidian geometry is not the only possible geometry, as was
believed in Kant's time, it should be possible to determine by means
of measurements whether, for instance, the sum of the three angles of
a triangle is 180 degrees or whether two straight parallel lines
always keep the same distance among them. If it were not the case,
then it would have been demonstrated experimentally that space is not
Euclidean. Thus the possibility of verifying the axioms of geometry
would prove that they are empirical and not given {\it a priori}.

Helmholtz developed his own version of a non-Euclidean geometry on
the basis of what he believed to be the fundamental condition for all
geometries: ``the possibility of figures moving without change of
form or size''; without this possibility, it would be impossible to
define what a measurement is. According to Helmholtz (1995, p. 244):
``the axioms of geometry are not concerned with space-relations only
but also at the same time with the mechanical deportment of solidest
bodies in motion.'' Nevertheless, he was aware that a strict Kantian
might argue that the rigidity of bodies is an {\it a priori}
property, but ``then we should have to maintain that the axioms of
geometry are not synthetic propositions... they would merely define
what qualities and deportment a body must have to be recognized as
rigid''.

At this point, it is worth noticing that Helmholtz's formulation of
geometry is a rudimentary version of what was later developed as the
theory of Lie groups, which I will mention in Section 3. As for the
transport of rigid bodies, it is well known nowadays that rigid
motion cannot be defined in the framework of the theory of
relativity: since there is no absolute simultaneity of events, it is
impossible to move all parts of a material body in a coordinated and
simultaneous way. What is defined as the length of a body depends on
the reference frame from where it is observed. Thus, it is
meaningless to invoke the rigidity of bodies as the basis of a
geometry that pretend to describe the real world; it is only in the
mathematical realm that the rigid displacement of a figure can be
defined in terms of what mathematicians call a {\it congruence}.

Arguments similar to those of Helmholtz were given by Reichenbach
(1958) in his intent to refute Kant's doctrine on the nature of space
and time. Essentially, the argument boils down to the following: Kant
assumed that the axioms of geometry are given {\it a priori} and he
only had classical geometry in mind, Einstein demonstrated that space
is not Euclidean and that this could be verified empirically, {\it
ergo} Kant was wrong.

However, Kant did not state that space must be Euclidean; instead, he
argued that it is a pure form of intuition. As such, space has no
{\it physical} reality of its own, and therefore it is meaningless to
ascribe physical properties to it. Actually, Kant never mentioned
Euclid directly in his work, but he did refer many times to the
physics of Newton, which is based on classical geometry. Kant had in
mind the axioms of this geometry which is a most powerful tool of
Newtonian mechanics. Actually, he did not even exclude the
possibility of other geometries, as can be seen in his early
speculations on the dimensionality of space (Kant 1986).

The important point missed by Reichenbach is that Riemannian geometry
is necessarily based on Euclidean geometry. More precisely, a
Riemannian space must be considered as {\it locally} Euclidean in
order to be able to define basic concepts such as distance and
parallel transport; this is achieved by defining a {\it flat} tangent
space at every point, and then extending all properties of this flat
space to the {\it globally} curved space (see, e.g., Eisenhart,
1959). To begin with, the structure of a Riemannian space is given by
its metric tensor $g_{\mu \nu}$ from which the (differential) length
is defined as $ds^2 = g_{\mu \nu}dx^{\mu}dx^{\nu}$; but this is
nothing less than a generalization of the usual Pythagoras theorem in
Euclidean space. As for the fundamental concept of parallel
transport, it is taken directly from its analogue in Euclidean space:
it refers to the transport of abstract (not material, as Helmholtz
believed) figures in such a space. Thus Riemann's geometry cannot be
free of synthetic {\it a priori} propositions because it is entirely
based upon concepts such as length and congruence taken form Euclid.
We may conclude that Euclid´s geometry is the condition of
possibility for a more general geometry, such as Riemann's, simply
because it is the natural geometry adapted to our understanding; Kant
would say that it is our form of grasping space intuitively. The
possibility of constructing abstract spaces does not refute Kant's
thesis; on the contrary, it reinforces it.

But then, can the axioms of geometry be verified experimentally? Let
us see what Poincar\'{e} had to say on that matter.

\section{Can geometry be an object of experience?}

This is the fundamental question put forward by Henri Poincar\'{e},
to which his answer was that something as a ``geometric experiment''
cannot be performed (Poincar\'{e}, 1952):

\begin{quote}
Think of a material circle, measure its radius and circumference, and
see if the ratio of the two lengths is equal to $\pi$. What have we
done? We have made an experiment on the properties of the matter with
which this {\it roundness} has been realized, and of which the
measure we used is made.
\end{quote}

Something similar would happen with astronomical observations. For
instance, Lobachevsky had suggested that it should be possible to
determine the curvature of the space we live in by measuring the
parallaxes of distant stars (Jammer, 1993, p. 149). But then
Poincar\'{e} pointed out:

\begin{quote}
What we call a straight line in astronomy is simply the path of a ray
of light. If, therefore, we were to discover negative parallaxes, or
to prove that all parallaxes are higher than a certain limit, we
should have a choice between two conclusions: we could give up
Euclidean geometry, or modify the laws of optics, and suppose that
light is not rigorously propagated in a straight line.
\end{quote}

He then concluded: ``It is needless to add that every one would look
upon this [second] solution as the more advantageous''. Poincar\'{e}
wrote his essay in 1898, when the utility of non-Euclidean geometries
in physics was still unknown. Evidently he was mistaken on this
point: fifteen years later Einstein showed that the geometry of
Riemann is more advantageous for the description of gravity.

Now the question is: why does Euclid's geometry seem so natural? Kant
assumed that the basic axioms of this geometry (the only geometry
known in his time) were given {\it a priori}, while Poincar\'{e} saw
the answer in the natural selection, thanks to which ``our spirit
{\it adapted} to the conditions of the outer world, adopted the {\it
most advantageous} geometry for the species; in other words, the most
{\it comfortable}\footnote{Underlined by H. P.}.'' There is no doubt
that this geometry is most comfortable since it was accepted as the
natural one over two millennia, but Poincar\'{e} never explained why
this was so. A Kantian, however, would argue that it is most
comfortable because it is given {\it a priori}. In any case,
Poincar\'{e}'s conclusion is clear: ``geometry is not true, it is
advantageous''. Accordingly, it is nowadays evident that Euclidean
geometry is most comfortable to use, except for the description of
the gravitational field, for which the Riemannian geometry is more
advantageous. But is Riemann's geometry indispensable for this task?
It is not, as I will show next.

\section{General relativity as a gauge theory}

With the general theory of relativity, Einstein successfully unified
physics with geometry. Space, combined with time in a single space of
four dimensions, was no longer a simple scenario for natural
phenomena, but acquired a fully dynamical role. Space-time, in
Einstein's theory, is a curved space and its curvature is equivalent
to a gravitational field. Thus non-Euclidean geometry found an
important application in physics. Furthermore, general relativity
made definite predictions, and it therefore appeared that the
non-Euclidean nature of space could be an empirical fact liable to be
verified.

What would  Poincar\'{e} have said on this point? Unfortunately, his
untimely death did not let him hear about the new theory of Einstein.
In order to defend his own point of view, he would surely have
insisted that the general theory of relativity simply implies that
light rays do not propagate along ``straight lines'' but in a more
complicated way. From the empirical point of view, it is perfectly
equivalent to saying that light moves along ``null geodesics'' in a
curved space, as postulated in the theory of relativity, or along a
curved path in Euclidean space: there is no way to distinguish by
experiments between the two possibilities, and therefore general
relativity does not contradict the conventionalism of geometry. In
fact, every prediction of general relativity can also be interpreted
as a phenomenon in a Euclidean space, since all the classical tests
can be ascribed to alterations of Newtonian physics. The perihelion
shift of Mercury could be due to a correction in Newton's law of
gravity; the deflection of light rays may be a real deflection
produced by gravity; and the gravitational red-shift of light is due
to an actual loss of energy. Even the formation of black holes does
not relay on the geometry of space: it is perfectly consistent to
assume that light is strongly bent around a massive compact object,
to the extent that it cannot escape from it.

As for cosmology, the expansion of the Universe is predicted by the
dynamical equations of Friedmann that follow from general relativity.
However, it is known from the work of McCrea and Milne (1934) that
these same equations can also be obtained within the framework of
Newtonian mechanics; it is only a matter of interpreting the
Newtonian variables in terms of their relativistic counterparts (see
Bondi, 1960, Chap. IX, for a detailed discussion). Moreover, it is a
noteworthy fact that present day observations are compatible with a
Universe that is spatially flat on the average; that is to say, light
rays do not diverge or converge, as would be the case in a space with
negative or positive curvature, but they move in ``straight'' paths
(actually, this supports Poincar\'{e}'s thesis even further: there is
no ``sufficient reason'', in a Leibnizian sense, for light rays to
either diverge or converge in a perfectly homogeneous and isotropic
universe).

In order to do justice to Poincar\'{e}, let us imagine for a moment
what would have been of physics if Einstein had not lived. Without
Einstein, special relativity would have been formulated anyhow, since
all the basic ingredients of this theory were known at the beginning
of the last century; it would have taken only an ingenious mind to
paste all the pieces together. However, it is undeniable that general
relativity is the sole achievement of Einstein, since it is a totally
original theory with no more antecedents than the classic physics of
Newton and the geometry developed in the XIX century.

What would have been, then, of general relativity without Einstein?
The answer can be found in the wider context of what is presently
known as a gauge theory, a concept that appeared in the fifties and
turned out to be extremely important in theoretical physics. With the
formalism of gauge theory, it is possible to deduce the equations
that describe an interaction on the sole basis of their symmetries;
in other words, symmetry determines dynamical laws! Gauge theories
are based on two fundamental mathematical concepts developed
throughout the XIX century: abstract mathematical spaces and the
theory of continuous groups of transformations. Mathematicians
discovered that it is possible to define abstract spaces without
resorting to any coordinate system, and that these more general
spaces have a much more complex and interesting structures than the
usual one of three dimensions. Curved spaces of many dimensions, such
as those of Riemann, are only particular cases. In quantum mechanics,
for example, the state of an atomic system is described as a vector
in Hilbert spaces: a Hilbert space can have any number of dimensions,
even infinite, and the ``coordinates'' are complex numbers. On the
other hand, the theory of continuous groups of transformations was
developed mainly by Sophus Lie. Continuous transformations in
abstract spaces generalize the concepts of rotations and translations
of material bodies according to rules of motion in usual space. The
crucial point is that a given generalized motion corresponds to a
particular symmetry invariance and can be described by a non
commutative algebra, forming a Lie group, which can be classified in
well defined categories.

The theory of Lie groups has crucial applications in physics,
particularly in the description of fundamental interactions between
atomic particles. The basic idea is that the various states of an
atomic system are described by vectors in an abstract space, and that
these vectors can be transformed without altering the physics of the
system, just as the translation of a solid body does not alter its
properties. The essential point is that these transformations can be
interpreted as symmetry properties and the mathematical formalism
permits one to deduce all the equations that describe the
interactions. Such a formalism was first used  by C. N. Yang and R.
L. Mills (1954) to obtain the equations that describe nuclear
interactions.

The idea put forward by Yang and Mills was generalized a year later
by Ryoyu Utiyama, in an article that had a great impact in
theoretical physics. Utiyama (1955) showed that if the symmetry
properties of a fundamental interaction are known in a given point of
an abstract space, then the dynamical equations of that interaction,
valid everywhere in the same space, could be deduced precisely.
Utiyama's formalism is based on the theory of continuous Lie groups.
Essentially, if the group of transformations that does not alter the
form of the interaction is known {\it locally}, then it is possible
to deduce the equations that are valid {\it globally}. More
specifically, Utiyama stated the problem in the following form:

\begin{quote}
Let us consider a system of fields $Q^A (x)$ which is invariant under
some transformation group $G$ depending on parameters $\epsilon_1$,
$\epsilon_2$, ...$\epsilon_n$. Suppose that the aforementioned
parameter-group is replaced by a wider group $G'$ derived by
replacing the parameters $\epsilon'$ by a set of arbitrary functions
$\epsilon'(x)$, and that the system considered is invariant under the
wider group $G'$.
\end{quote}

Under these conditions, he showed that a new field $A(x)$ can be
introduced and the new Lagrangian $L'(Q,A)$ can be deduced from the
original one $L(Q)$, together with the field equations. The idea is
that, given a Lie group with operators $\widehat{T}_a$ that satisfy
the commutation relations $[\widehat{T}_a , \widehat{T}_b ] =
f^c_{ab} \widehat{T}_c$, the derivative of a field $Q^A$ must be
generalized to $\partial_{\mu} Q^A \rightarrow \partial_{\mu} Q^A -
T^A_{a~B} Q^B A^a_{\mu}$, with a new field $A^a_{\alpha}$ that in
turn defines a gauge field
\begin{equation}
F^a_{\mu\nu}= \partial_\mu  A^a_{\nu} - \partial_\nu A^a_{\mu} -
\frac{1}{2} f^a_{bc} ( A^b_{\mu} A^c_{\nu} - A^b_{\nu}
A^c_{\mu})~,\label{F}
\end{equation}
in terms of the structure constants of the Lie group, $f^a_{bc}$.

Gauge field theory has found its paramount application in the
Standard Model of elementary particles, a model that has been most
successful in describing nuclear and electromagnetic interactions.
Actually, the model is the gauge theory associated to a combination
of three simple groups of transformations in an abstract space:
SU(3), SU(2) and U(1).

As for the general theory of relativity, Utiyama showed in his
article of 1956 that it is actually the gauge theory associated to
the group of Lorentz transformations in Minkowski space. In fact, it
is clearly seen from the formulation of Utiyama that there is an
equivalence between the electromagnetic and the gravitational fields:
the electromagnetic field is a spin-1 field described by a tensor of
rank 2, and the gravitation field is a spin-2 field described by a
tensor of rank 4. The gravitational field tensor, by a lucky
coincidence, has precisely the same algebraic structure as the
Riemann tensor that characterizes a Riemannian space of four
dimensions.

Let us paraphrase Utiyama's formulation in a more modern notation
using the language of differential geometry. The formulation runs
along the following lines. Given the Minkowski metric $\eta_{ab}=(-1,
1,1,1)$ in flat space with Cartesian coordinates, define a tetrad
(also known as {\it Cartan rep\`ere mobile} or {\it vierbein})
$e^a_{\alpha}$ such that, at a given point, the differential line
element is\footnote{Utiyama used the notation $h^a_{\alpha}$ for
$e^a_{\alpha}$, but he did not recognize this as a tetrad.}
\begin{equation}
ds^2=  \eta_{ab} e^a e^b = g_{\alpha\beta}dx^{\alpha}dx^{\beta},
\end{equation}
where $g_{\alpha\beta}$ is the metric tensor and
$e^a=e^a_{\alpha}dx^{\alpha}$ is a one-form (see, e.g., Flanders
1963). Under an infinitesimal Lorentz group transformation, the
tetrad $e^a$ transforms as $e^a \rightarrow  e^a + \epsilon^a_{~~b}
~e^a$, where $\epsilon_{ab} =- \epsilon_{ba}$. Next, let $Q^A$ be a
tensor with $A$ being an index refereing to a particular irreducible
representation of the Lorentz group. Then, for a Lagrangian $L(Q^A,
\partial_{\mu}Q^A)$ to be invariant under a ``generalized Lorentz
transformation'' with $\epsilon_{ab}(x)$, the partial derivatives
must be substituted by
\begin{equation}
\partial_{\mu}Q^A \rightarrow \partial_{\mu}Q^A - T_{ab~~B}^{~~A} Q^B A^{ab}_{\mu}~,
\end{equation}
where $\widehat{T}_{ab}$ are the generators of the Lorentz group,
\begin{equation}
[\widehat{T}_{ab},\widehat{T}_{cd}] = \frac{1}{2}
f_{ab~~~cd}^{~~mn}~\widehat{T}_{mn}.
\end{equation}
Then, according to (\ref{F}), a new field is defined as
\begin{equation}
R^{ab}_{~~\mu\nu}= \partial_\mu  A^{ab}_{\nu} - \partial_\nu
A^{ab}_{\mu} - \frac{1}{2} f^{~~ab}_{kl~~~mn} ( A^{kl}_{\mu}
A^{mn}_{\nu} - A^{mn}_{\mu} A^{kl}_{\nu}).
\end{equation}

Thus the field is described by the tensor $R^{ab}_{~~\mu\nu}$ from
which the scalar $R= e_a^{\mu} e_b^{\nu}R^{ab}_{~~\mu\nu}$ can be
constructed. The next step is to take ${\rm det}(e^a_{\alpha}) R$ as
the Lagrangian of the field, add to it the Lagrangian of the
matter-source, and, as explained in any text book of general
relativity, obtain the usual Einstein equations by varying with
respect to $g_{\alpha \beta}$.

Now, in the language of differential geometry, if $e^a_{\alpha}$ is
identified with a tetrad, then $A^{ab}_{\alpha}$ are the Ricci
rotation coefficients (related to the Christoffel symbols) and
$R^{\alpha\beta}_{~~\mu\nu}=e^{\alpha}_a e^{\beta}_b
R^{ab}_{~~\mu\nu}$ is the Riemann tensor (see, e.g., Flanders 1963).

Just for the sake of comparison, recall that the electromagnetic
field $F_{\alpha\beta}$ can be deduced as the gauge field of the
Abelian group $U(1)$: $F_{\alpha\beta}=\partial_{\alpha} A_{\beta} -
\partial_{\beta} A_{\alpha}$, where the electromagnetic potential
$A_{\alpha}$ plays the same role as the Ricci rotation coefficients
for the gravitational field. The Lagrangian is taken as
$L=\frac{1}{4\pi} F_{\alpha\beta}F^{\alpha\beta}$, which is quadratic
in the field; unlike the case with the Riemann tensor, it is not
possible to construct a scalar from $F_{\alpha\beta}$ that is linear
on the field; this is an important difference between the two fields.

As for the equations of motion for a test particle, these can be
obtained, in the case of an electromagnetic field, from the action
$\int A_{\mu} U^{\mu} d\tau$ where $U^{\mu}$ is the four-velocity of
the particle and $\tau$ its proper time: the Lorentz force equation
follows. For the gravitational field, the action is of the form $
\int \sqrt{g_{\mu \nu} U^{\mu} U^{\nu}} d\tau$, and its variation
gives rise to the well known geodesic equations {\it if} $g_{\mu
\nu}$ is interpreted as the metric of a Riemannian space.

Looking the matter from a modern perspective, it can be ascertained
now that Einstein used an extremely ingenious logical reasoning in
order to develop a theory based on the {\it formal} analogy between
the equations of Riemannian geometry and those of a relativistic
gravitational field. Thus Einstein was four decades ahead of his time
in formulating what would be known as a gauge theory. If Einstein had
not had this insight, it would certainly correspond to Utiyama the
merit of formulating a fully relativistic theory of gravity.

The formulation of a gravitational theory by Utiyama does not relay
on a curved space, but rather on an abstract space that has the same
mathematical structure as a Riemannian space of four dimensions. In
this way he arrived  to a basic set of equations that are identical
to those obtained by Einstein. Thus, Poincar\'{e} would be perfectly
right in claiming that it is a mere question of convenience whether
to use one formulation or the other: there is no difference, from a
purely formal point of view, to use Riemannian geometry or gauge
field theory, although the elegant formulation of Einstein is more
comfortable and easier to visualize because it is based on a
beautiful geometric analogy.

\section{An imaginary dialogue}

Poincar\'{e} and Einstein met only once, during the 1911 Solvay
Congress at Brussels, a year before the untimely death of the great
French mathematician. Poincar\'{e} already knew about the special
theory of relativity, but he was not entirely convinced of it,
although he had a great esteem for his young creator. As for
Einstein, who was usually reluctant to acknowledge the contributions
of his predecessors, it is only on his later years that he mentioned
Poincar\'{e} as he deserved.

In 1949, Paul Schilpp edited a volume dedicated to Einstein with
several essays written by distinguished scientists and philosophers
of science (Schilpp 1949). In the final chapter, Einstein commented
each essay and, in particular, he took the opportunity to imagine a
dialogue between Poincar\'{e} and Reichenbach on the geometric
conventionalism that the latter author had criticized in his essay.
Einstein summarized the antagonist positions in a simple and clear
way: ``Is a geometry ---looked at from the physical point of view---
verifiable (viz., falsifiable) or not? Reichenbach, together with
Helmholtz, says: Yes, provided that the empirically given solid body
realizes the concept of `distance'. Poincar\'{e} says no and
consequently is condemned by Reichenbach ''.

Next, Einstein imagined the following dialogue:

\begin{quote}
{\it Poincar\'{e}}: The empirically given bodies are not rigid, and
consequently cannot be used  for the embodiment of geometric
intervals. Therefore, the theorems of geometry are not verifiable.

{\it Reichenbach}: I admit that there are no bodies which can be
credited {\it immediately} adduced for the ``real definition'' of the
interval. Nevertheless, this real definition can be achieved by
taking the thermal volume-dependence, elasticity,  electro- and
magneto-striction, etc., into consideration. That this is really and
without contradiction possible, classical physics has surely
demonstrated.

{\it Poincar\'{e}}: In gaining the real definition improved by
yourself you have made use of physical laws, the formulation of which
presupposes (in this case) Euclidean geometry. The verifications, of
which you have spoken, refer, therefore, not merely to geometry but
to the entire system of physical laws which constitute its
foundation\footnote{Underlined by A. E.}.
\end{quote}

At the end of the imaginary conversation, Einstein manifested his
agreement with Kant in so far as ``there are concepts (as, for
example, that of causal connection), which play a dominating role in
our thinking, and which, nevertheless, cannot be deduced by means of
a logical process from the empirically given.'' According to
Einstein, this was Kant's most important contribution and not the
belief that ``Euclidean geometry is necessary to thinking and offers
{\it assured} (i.e., not dependent upon sensory experience) knowledge
concerning the objects of `external' perception\footnote{Underlined
by A. E.}''.

It is thus evident that Einstein was still convinced of the
non-Euclidean nature of space, but he was unable to give a convincing
argument against Poincar\'{e} and only paraphrased him. There is no
doubt that Einstein believed that the curvature of space must be an
object of experience; nevertheless, it can be seen from the above
dialogue he imagined, that he was quite aware that the problem is
considerably more difficult than what Helmholtz or Reichenbach
thought. Surely, his conviction was sustained on the fact that,
during his lifetime, the only known form of unifying gravity and
relativity was using the mathematical tool of Riemannian geometry.
The development of gauge theories and, particularly, the independent
formulation of Utiyama of the same relativistic theory of gravity
(which Einstein did not live to see), made it evident that Riemannian
geometry is very advantageous for the theory describing this
fundamental interaction, but, as Poincar\'{e} would have said, it is
a very convenient convention, but a convention anyway.

\vskip1cm

The author is grateful to M. P. Ryan for stimulating discussions.

\newpage

\section*{References}

{\noindent Bondi, H. (1960). {\it Cosmology}. Cambridge: Cambridge
University Press.}
 \vskip0.3cm

{\noindent Eisenhart, L. P. (1959). {\it Riemannian Geometry}.
Princeton: Princeton University Press.} \vskip0.3cm

{\noindent Flanders, H. (1963). {\it Differential Geometry}.  San
Francisco: Academic Press.}

\vskip0.3cm

{\noindent Helmholtz, H. von (1995). On the Origin and Significance
of Geometrical Axioms. In {\it Science and Culture}. Chicago: Univ.
Chicago Press.}

\vskip0.3cm

{\noindent Jammer, M. (1993). {\it Concepts of Space}. New York:
Dover.}

\vskip0.3cm

{\noindent Kant, I. (1929). {\it Critique of Pure Reason}. Boston/New
York: MacMillan \& Co., Bedford/St. Martin's.}

\vskip0.3cm

{\noindent Kant, I. (1986), {\it Gedanken von der wahren
Sch$\ddot{a}$tzung der lebendigen Kr$\ddot{a}$fte}, ({\it Thoughts on
the True Estimation of Living Forces}). Amsterdam: Editions Rodopi.}

\vskip0.3cm

{\noindent McCrea, W. H. and Milne, E. A. (1934). Newtonian universes
and the curvature of space. {\it Quart. J. Math. (Oxford Series), 5,
73-80}

\vskip0.3cm

{\noindent Poincar\'{e}, Henri (1952). Experiment and Geometry. In
{\it Science and Hypothesis}. New York: Dover.}

\vskip0.3cm

{\noindent Reichenbach H. (1958). {\it The Philosophy of Space and
Time}. New York: Dover}

\vskip0.3cm

{\noindent Schilpp, P. A., editor (1949). {\it Albert Einstein:
Philosopher-Scientist}. Cambridge: Cambridge Univ. Press.}

\vskip0.3cm

{\noindent Utiyama, R. (1956). Invariant Theoretical Interpretation
of Interaction. {\it Physical Review, 161, 1597-1607.}

\vskip0.3cm

{\noindent Yang, C. N. and Mills, R. L. (1954). Conservation of
Isotopic Spin and Isotopic Gauge Invariance. {\it Physical Review,
96, 191-195.}

\end{document}